\documentclass[12pt,preprint]{aastex}
\begin{document}

\title{Correlated Depletion and Dilution of Lithium and Beryllium Revealed 
by Subgiants in M 67}

\author{Ann Merchant Boesgaard\altaffilmark{1}, Michael
G.~Lum\altaffilmark{1}} 

\affil{Institute for Astronomy, University of Hawai`i at M\-anoa, \\ 2680
Woodlawn Drive, Honolulu, HI {\ \ }96822 \\ } 

\email{boes@ifa.hawaii.edu}
\email{mikelum@ifa.hawaii.edu}

\author{Constantine~P.~Deliyannis\altaffilmark{1}}
\affil{Department of Astronomy, Indiana University, 727 East 3rd Street, \\
Swain Hall West 319, Bloomington, IN {\ \ }47405-7105 \\ }

\email{cdeliyan@indiana.edu}

\altaffiltext{1}{Visiting Astronomer, W.~M.~Keck Observatory jointly operated
 by the California Institute of Technology and the University of California.}

\begin{abstract}
The surface content of lithium (Li) and beryllium (Be) provide
insights into the mixing and circulation mechanisms in stellar
interiors.  The old open cluster, M 67, has been well-studied for Li
abundances in both main-sequence and evolved stars.  The Be abundances
give us a probe to a deeper level in stars.  We have taken
high-resolution spectra with Keck I with HIRES to determine Be
abundances along the subgiant branch of M 67, where there are dramatic
depletions of Li.  These subgiants range in mass from 1.26 to 1.32
M$_{\odot}$ and have evolved from main-sequence stars that would have
occupied the region of the Li-Be dip found in younger clusters.
Lithium abundances have been adjusted to the same scale for 103 stars
in M 67 by Pace et al.  The more massive stars $-$ now the coolest and
furthest-evolved from the main-sequence $-$ show a drop in Li by a
factor of 400 across the subgiant branch.  Our new Be abundances also
show a decline, but by a factor of $\sim$50.  The two elements decline
together with Li showing a steeper decline in these subgiants than it
does in the Li-Be dip stars.  The relative decline in Be abundance
compared to Li is remarkably well fit by the models of Sills \&
Deliyannis, made specifically for the subgiants in M 67.  Those models
include the effects of mixing induced by stellar rotation.  These M 67
subgiants show the effects of {\it both} main-sequence depletion and
post-main-sequence dilution of both Li and Be.
\end{abstract}


\section{INTRODUCTION}

The surface abundances of the rare light elements, lithium (Li),
beryllium (Be), and boron (B), allow us to probe into the internal
structure of the stars.  This results from the fact that these three
elements are readily destroyed in stellar interiors where the
temperatures are only a few million degrees Kelvin.  They also provide
insights into the mixing and other physical mechanisms taking place
there.

Figure 1 -- a representation of the Sun -- indicates why Li and Be
provide such important information.  The Li atoms will be destroyed by
nuclear reactions when they encounter a temperature of $\sim$2.5 x
10$^6$ K and higher so Li exists only in the outer region of a star.
(Above the red circle in Figure 1.)  The temperature has to reach
$\sim$3.5 x 10$^6$ K to destroy Be atoms by nuclear reactions so the
surface reservoir containing Be is larger than that for Li.  (Above
the blue circle in Figure 1.)  The most robust of these three light
elements is B; those nuclei need to be mixed down to $\sim$5 x
10$^6$ K.  (Not shown in this figure.)

All three light elements are susceptible to destruction in stellar
interiors by nuclear reactions; some of these reactions are of the
type (p,$\alpha$), (p,$\gamma$), ($\alpha$,n), ($\alpha$,$\gamma$)
(Fowler et al.~1975).  The major reaction for Li is
$^{7}$Li(p,$\alpha$)$^{4}$He showing that when the $^{7}$Li nucleus
encounters a proton it will become 2 $^4$He nuclei (= 2 alpha
particles) at $T \sim$ 2.5 x 10$^6$ K.  If the measured stellar
photospheric abundances of Li, Be, B are less than the initial values
(taken to be meteoritic abundances), then we know some destruction has
occurred. (We measure these same initial values in young clusters,
such as the Hyades (Boesgaard et al.~2016).)  For example, if all of
the surface Li is absent but some of the Be is present and all of the
B is present, we know how deep the mixing has gone inside the star.
The process of the mixing and destruction is a main sequence
phenomenon and is very slow.  We know this by comparing amounts of Li
and Be in stars and clusters of an array of ages.

Observations of both Li and Be show that these elements are subject to
{\it depletion} in main-sequence stars indicating destruction by
nuclear reactions.  The surface abundances are depleted when the
surface convection zone (SCZ) is deep enough so that its base is hot
enough to destroy these elements.  (We note that in standard theory
i.e.~no rotation, magnetic fields, diffusion, mass loss or gain,
etc.~, Li is only depleted during pre-main-sequence evolution for G
dwarfs and hotter stars, see Deliyannis et al. 1990)

As stars evolve off the main-sequence, especially in the red giant
phase, Li and Be are subject to {\it dilution}.  This means that the
surface regions, which are relatively rich in Li and Be, are mixed
with deeper regions where there is no Li or Be.  This mixing is
primarily due to the deepening of the outer convection zone as the
star expands.  The phenomenon of dilution was discussed in giants by
Iben (1965, 1967) based on observations by Wallerstein (1965, 1966).

Boesgaard \& Tripicco (1986) discovered a discontinuity in the Li
abundances in main-sequence stars in the Hyades.  (This Li-dip was
presaged some 20 years earlier by Wallerstein et al.~(1965) who found
that they could determine only upper limits on Li for seven of their
Hyades stars with $(B-V)$ values near 0.45.)  There are large
depletions of Li in a narrow temperature range of 6400 - 6800 K
(masses near 1.2 $M_{\odot}$).  The Li abundances in stars on either
side of this ``Li dip'' are normal and close to the Li content
measured in meteorites (Lodders 2003).  Boesgaard \& Budge (1988)
further delineated the sharp drop in the mid-F dwarfs and Boesgaard et
al.~(2016) showed the total Li-temperature profile for the Hyades.
This Li-dip is not present in the younger Pleiades cluster
(Pilachowski et al.~1987, Boesgaard et al.~1988). This indicates that
the Li depletion occurs during main-sequence evolution, rather than
during the pre-main-sequence phase.  Additional observations by
Boesgaard and collaborators and other groups found this phenomenon to
exist in F dwarfs in other older open clusters and in field stars.
The Li-dip strongly contradicted the strictures of the standard
theory, and suggested that additional physical mechanism(s) must be
acting inside stars during the main sequence.  This realization
inspired exploration of a number of such mechanisms, including mass
loss, diffusion, and slow mixing induced by rotation or gravity waves
(discussion in Deliyannis et al. 1998).

\section{BERYLLIUM IN CLUSTERS}

It is important to have abundance information on {\it both} Li and Be
to probe different depths in a star.  This is clearly shown in our
results for both Li and Be in the Hyades cluster in Figure 2 derived
from Boesgaard et al.~(2016).  The abundances are plotted on the same
vertical scale and normalized to their relative initial abundances.
We use the notation, A(Li) = log N(Li)/N(H) +12.00 and A(Be) = log
N(Be)/N(H) +12.00.  The Li dip in the F stars ($T_{\rm eff}$
$\sim$6400-6800 K) is very deep.  While there is a dip in the Be
abundance in that temperature range, it is not as dramatic.  While
A(Li) drops by over two orders of magnitude, A(Be) falls by a factor
of a little over six.  As Figure 1 shows, the Be nuclei have to be
mixed to deeper layers to be destroyed.  In the cooler stars ($<$6200
K) there is a systematic drop in Li, but this is not accompanied by a
drop in Be.  The surface content of these two atoms provide
information about which processes cause mixing leading to depletions
at different levels in the interiors of stars of different
temperatures.

Boesgaard \& King (2002) found the dip in the Be abundances in the Hyades,
similar to that of Li, but Be is not as severely depleted as Li.  Other young
open clusters were studied for Be in the-Li dip region as well e.g.~Pleiades
and $\alpha$ Per (Boesgaard et al.~2003a), Coma and UMa (Boesgaard et
al.~2003b), Praesepe (Boesgaard et al.~2004a).  They found no Be dip in the
young Pleiades and relatively small Be-dips in the older clusters.  

The various mechanisms proposed to create the Li-dip predict differing
effects on Be, so Be data are critical to help distinguish among them.
Discovery of a correlated depletion between Li and Be supported
rotationally-induced mixing as the dominant mechanism creating the
Li-Be dip.  It could not support diffusion and mass loss, and even
favored slow mixing due to rotation over slow mixing to gravity waves
(Deliyannis et al. 1998).  The correlated depletion of Li and Be was
further detailed by Boesgaard et al.~(2004b) in a sample of 88
main-sequence stars in clusters and the field.  For the 35 stars on
the cool side of the dip with temperatures of 6300 - 6650 K, the slope
of the relationship A(Be) vs.~A(Li) was found to be +0.43.

A(Be) = 0.43($\pm$0.04)A(Li) - 0.17($\pm$0.09)

The models of Charbonnel (1994) and Deliyannis \& Pinsonneault (1997)
of rotationally-induced stellar mixing predict 1) depletion of both Li
and Be in the Li-Be dip, 2) greater depletion of Li than Be, and 3)
are consistent with this slope.  It should be noted that a small B-dip
has been discovered in field stars, suggesting that mixing extends to
deep enough layers to affect the surface B abundances, consistent with
models that include rotational mixing (Boesgaard et al. 2005, 2016).

\section{M 67 = NGC 2682}

The old, open cluster, M 67, is similar to the Sun in metalicity and
age.  The best fitting isochrone in VandenBerg et al.~(2007) gives an
age of 3.9 Gyr.  Yadav et al.~(2008) compare their photometric data
with four different libraries of stellar models. They find a range in
age of 3.5 -- 4 Gyr and prefer 3.8 Gyr.  The metallicity M 67 has been
found in many studies.  Those for main-sequence stars include Friel \&
Boesgaard (1992) with [Fe/H] = +0.02; Randich et al.~(2006) with
+0.03; Pace et al.~(2008) with +0.03; Jacobsen et al.~(2011) with
$-$0.01; Canto Martins et al.~(2011) with $-$0.05; \"Onehag et
al.~(2014) with $-$0.01; Souto et al.~(2018) with $-$0.03; Lum (2018)
with $-$0.02.

This cluster provides uniquely important insights about the physical
origin of the Li-Be-B dip because its subgiants are evolving out of
the middle of the Li-dip.  As they evolve and their SCZs deepen, they
reveal the profile of the Li and Be preservation regions (i.e.~Li and
Be abundances as a function of depth), which varies from mechanism to
mechanism. Furthermore, M 67 subgiants have the potential to probe a
larger range of Li and Be (and B) abundances than has been possible
for main sequence stars.  Thus they provide even more insightful
information and stringent constraints.  Finally, it is possible that
by the time Li-dip stars evolve to the much more advanced age of M 67
(as compared to the Hyades), mechanisms other than rotational mixing
might also become important. Sills \& Deliyannis (2000) argued that Li
abundances in M 67 subgiants favored rotational mixing, but it was
possible that diffusion and mass loss might play a role.  We once
again turn to the critically important element Be, in order to
determine the relative importance of these various mechanisms (or to
suggest some other ones).

Early studies of Li in M 67 were done by Hobbs \& Pilachowski (1986)
of eight stars including one subgiant, by Spite et al.~(1987) of six
main-sequence stars, and by Garcia L\'opez et al.~(1988) of eight
stars including one giant.  A large spread in A(Li) was found.  This
was followed by Li abundance determinations by Balachandran (1995),
Pasquini et al.~(1997), Jones et al.~(1999).  More recently Randich et
al.~(2007) looked at both Li and Be in four main-sequence stars, two
turn-off stars and two blue stragglers.  In their search for solar
twins in the M 67 Pasquini et al.~(2008) determined Li abundances in
59 single stars in M 67 and identified 10 solar twins.  (Three of
these were studied in more detail by Castro et al.~(2011).)  In a
later study Canto Martins et al.~(2011) found Li abundances in 14
turn-off stars and subgiants and in 13 giants.

The Li abundances for 103 stars in M 67 were normalized to the same
temperature scale by Pace et al.~(2012) with consistent stellar
parameters.  These are the Li abundances and stellar parameter
determinations we use in this work on Be abundances in our nine
subgiants.

\section{SPECTROSCOPIC OBSERVATIONS AND DATA REDUCTION}

We have observed nine subgiants in M 67 for which Li abundances had
been previously determined.  Figure 3 shows the color-magnitude
diagram for M 67 with photometry from Montgomery et al.~(1993).  The
horizontal and vertical lines enclose the region of our target stars.
The ones we have observed for Be are shown surrounded by red squares.

The solar system/meteoritic abundance of Be is very low: log N(Be/H) =
1.41, where log N(H) is 12.00 (Lodders 2003).  This indicates that
stellar Be abundances need to be measured in the strong resonance
lines, typically Be II at $\lambda$3130 and $\lambda$3131.  Our
spectra were obtained with the Keck I telescope and HIRES (Vogt et
al.~1993).  The 2004 upgraded version of
HIRES\footnote{\url{http://www.ucolick.org/$\sim$vogt/hires\_2004ccd.html}}
has a detector with three CCDs and the blue chip has a quantum
efficiency of 93\% (!) at the wavelength of 3130 \AA{} where the Be II
lines are found.  This UV response is extremely important for our Be
work.  Our spectra have high spectral resolution ($\sim$45,000) with a
linear scale of 0.023 \AA{} pix$^{-1}$ and signal-to-noise ratios
(S/N) at 3131 \AA{} of 40-50.  The spectral range is $\sim$3035 --
5880 \AA.

We had three observing runs that covered six nights for this and other
projects.  Our individual exposure times were usually 30 minutes and
multiple exposures were taken each night of a given star with some
additional exposures on other observing nights.  Our program stars had
V magnitudes of 12-13.  We were aiming to obtain S/N near 50 per
pixel.  The exposure times were meant to be long enough to obtain good
enough signal to combine exposures, while short enough to minimize the
occurrence of cosmic-ray events impacting the detector.  Our spectra
were taken when each star was as close to the meridian as possible to
lessen the effect of atmospheric dispersion and absorption, which
especially affect the shortest wavelengths.  The log of the
observations is given in Table 1.  The names of the stars are from
several sources.  The F designation is from the early work of
Fagerholm (1906); Sand is from Sanders (1977); MMJ is from the
photometric work of Montgomery et al.~(1993); YBP refers to Yadav
et al.~(2008).  The exposure times are the total of multiple exposures
for a given night.  The S/N ratios are the combined values for stars
observed on more than one observing night.

In order to do the data reduction, we obtained two 1 s exposures of
the Th-Ar lamp at the beginning of the night and one more at the end
of the night.  Several exposures were taken of the quartz lamp for the
flat-fielding of the science frames.  The exposure times were 1 s for
the red CCD chip, 3 s for the green chip and 50 s for the blue chip.
(The long exposures on the blue chip were needed to get enough signal
in the shortest wavelengths for the Be II spectral lines.)  In
addition, at least 11 bias frames were taken at 0 s for the
calibration.

The MAKEE pipeline (Barlow
2008\footnote{\url{http://www.caltech.edu/$\sim$tb/makee/}}) was used
to subtract the master bias frame (from our 11+ bias exposures) and to
normalize the spectra with our master flat fields made for each of the
three CCD chips.  With the Th-Ar spectra (which were identical from
the beginning and the end of the night) we made the preliminary
wavelength justification.  We used IRAF\footnote{IRAF is distributed
by the National Optical Astronomy Observatories, which are operated by
The Association of Universities for Research in Astronomy, Inc., under
cooperative agreement with the National Science Foundation.} for the
final wavelength correction, for co-adding the multiple exposures of
each star and for the continuum fitting.

An example of the spectrum of F127 = YBP963 in the region of the Be II
lines is shown in Figure 4.  Some of the other strong lines in the
region are identified from the solar spectrum, but we note that all
the lines are blends with other atomic and molecular features.
Examples of the spectra for three of the stars are shown in Figure 5
where the positions of the Be II resonance are shown.  These three
stars cover the range in $T_{\rm eff}$ of the stars in our sample and
in A(Li) as computed by Pace et al.~(2012).

\section{ABUNDANCE DETERMINATION}

The abundance of Li has been determined in main-sequence, subgiant and
red giant stars in M 67 by several different groups.  They have been
put on the same temperature scale by Pace et al.~(2012).  The 103
stars all have masses $>$0.9 M$_{\sun}$.  (The individual references
and calibrations are given in the Pace et al.~paper.)  We have used
the values they give in their Table 1 for $T_{\rm eff}$ and log g.
The value for [Fe/H] = +0.02 was used in the models.  (We also used
[Fe/H] = $-$0.04 in our models, but found a mean difference for the
nine stars of only $-$0.004 with the lower Fe abundance.)  Edvardsson
et al.~(1993) determined an empirical relationship for microturbulent
velocity, $\xi$, from 189 F and G disk dwarfs with dependencies on
both $T_{\rm eff}$ and log g.  We have used that relationship to find
the appropriate value for each of our M 67 stars.  The stellar
parameters are given in Table 2.

We used the {\it synth} driver in the updated version of
MOOG\footnote{\url{http://www.as.utexas.edu/$\sim$chris/moog.html}}
(Sneden 1973, Sneden et al.~2012).  This version of MOOG includes
Kurucz's UV opacity edges, the metal bound-free opacity contributions
adapted from the ATLAS model atmosphere code (e.g.~Kurucz et al.~(2011
and references therein).  Our line list covers the region from 3129.5
to 3132.5 and has over 300 atomic and molecular lines.  Figure 6 shows
the observed and synthesized spectra for four of our stars.  The
notation is A(Be) = log N(Be)/N(H) + 12.00.  The abundance results for
A(Li) and A(Be) are given in Table 3 along with the stellar
temperature and mass.

We have tried to make estimates of the errors in our Be determinations
by using Kurucz grid models over our range in stellar parameters.  The
Be abundances from Be II are quite insensitive to temperature in the
region from 6750 to 5500 K.  For an error in temperature of $\pm$80 K,
the error in A(Be) is $\pm$0.01.  However, the Be abundance is
sensitive to log g.  For an error in log g of $\pm$0.25, the error in
A(Be) is $\pm$0.09 at T = 6500 K; $\pm$0.11 at T = 6250 K; $\pm$0.125
at T = 6000; and $\pm$0.13 at T = 5750 K.  As mentioned above we found
that the values we used for [Fe/H] between $-$0.04 to +0.02 had
virtually no effect on the Be abundance.  Another source of
uncertainty is how well the synthetic spectrum matches the observed
one.  In the determination of the best fit we relied almost completely
on the relatively unblended Be II line at 3131.067 line.  In Figure 6
we show the best fit Be abundance and those that are a factor of two
($\pm$0.3 in the log) above and below the best fit.  In the region of
that less blended line, we found sigma of the observed minus computed
spectra.  That led to an estimated error of $<$0.06 dex.  The log g
error is the dominant source of uncertainty in the Be abundance
determination.  The values of log g used by Pace et al.~(2012) were
computed with the isochrone determined by Castro et al.~(2011).  They
report values to three decimal points, but do not state the error.  We
will assume a typical error for log g of $\pm$0.25.  We adopt a
typical error in A(Be) of $\pm$0.14.

\section{RESULTS AND DISCUSSION}

The examination of both Li and Be abundances in field stars and
cluster stars has been important in determining the cause(s) of the
observed depletions.  Deliyannis \& Pinsonneault (1997) were able to
rule out mass loss and microscopic diffusion and show that the
timescale for mixing had to be long.  With new Be observations in some
60 field stars Stephens et al.~(1997) were able to reject those
mechanisms also and to show that neither meridional circulation nor
gravity waves could reproduce mild to severe Be depletions.
Deliyannis et al.~(1998) showed that the Li and Be abundances
(depletions) were correlated.  This pointed to rotationally-induced
mixing as the cause of the element depletions.  As explained by
Stephens et al.~(1997), the stellar wind removes angular momentum and
so an angular velocity gradient develops between the outer layers and
the rapidly-rotating interior.  In turn, a shear instability may
develop and will redistribute angular momentum and material through
the inner regions of the star.  This material mixing will slowly
deplete Li and Be, but to different degrees because Be is less
vulnerable to depletion, i.e.~a higher temperature is needed for the
nuclear reactions that destroy Be nuclei.

The stars we have observed for Be in the subgiants cover a large range
in A(Li) of nearly 3 orders of magnitude and also a large range in
A(Be) of nearly two orders of magnitude.  Figure 7 shows the Li
results from Pace et al.~(2012) as a function of their values for the
stellar mass.  (We have omitted the stars they call ``deviant'' from
this plot.)  In this figure we show a delineation by mass of
main-sequence stars (MS), subgiant stars (SG) and giant stars (RGB)
with the vertical dotted lines.  We take the turn-off mass as 1.26
M$_{\odot}$ from Sills \& Deliyannis (2000).  The points circled in
red are the subgiants for which we have taken spectra in the Be II
region.  These stars cover the full range in A(Li).  The more massive
subgiants are the ones that have evolved farthest along the subgiant
branch.  One can see that A(Li) plummets with increased mass.  As
Figure 1 shows, Li is far more susceptible to destruction than Be
because Li is destroyed at higher-layers (cooler-temperatures) in the
interiors of stars.

These subgiants have evolved from the temperature/stellar mass values
which show the strong Li dip in younger open clusters.  This can be seen
in Figure 8 which compares the Li abundances in the much younger
Hyades main-sequence stars with the Li in the main-sequence stars of M
67.  There are no stars left on the M 67 main sequence from the
temperature and mass range of the Li-Be dip.  So the M 67 subgiants
are the ones that once resided in the Li dip region during their
lifetime on the main sequence.  

Note that both Figures 7 and 8 show that there is a large range in
A(Li) for the solar-mass (Figure 7) and solar-temperature (Figure 8)
dwarfs in M 67.  The cooler dwarfs in M67 have a much greater spread
in A(Li) than that found in the Hyades.  The position of the Sun is
shown in both figures.  The value of the solar photospheric Li
abundance is similar to that in the M 67 dwarfs of the same mass and
temperature.  The values range from A(Li) = 0.6 to 1.8.  This
indicates that the Sun, which is similar in age and metallicity to M
67, does not have an anomalous Li value at A(Li) = 1.05 $\pm$0.10
(Asplund et al.~2009) with respect to M 67 solar=mass stars.

Although M 67 is about 6 times older than the Hyades, we can make a
comparison between the two clusters while bearing in mind their age
difference.  In fact, we can see that Li depletion continues on the
cool side of the Li-dip in a cluster three times the age of the
Hyades, NGC 6819 (Deliyannis et al.~2019).  Presumably, the M 67
cluster would have had a similar distribution and depletion of Li and
Be when the present subgiants were still on the main sequence.  The
added years for M 67 stars would result in even greater depletion of
Li and Be than found in the Hyades.  Figure 2 showed the distribution
of both Li and Be in the Hyades.  Although there is also a dip in the
Be abundances in the Li dip region, it is not as deep as the Li dip.
All the stars in the Li-Be dip have Be detections, not upper limit
results, whereas deep depletions and upper limits are seen for Li.  In
M 67 the subgiants would have evolved from stars with large Be
depletions and even larger depletions of Li.

It is possible that the Li and Be abundances in the subgiants of M 67
are completely consistent with their evolution from the Li-Be dip
region.  Inasmuch as M 67 is older than the Hyades, these stars would
have been on the main sequence, depleting Li and Be, for a longer
time.  However, those main-sequence (surface) abundances in M 67 stars
would start to be diluted by the expansion of the surface convection
zone as the stars evolve off the main-sequence.

We show the Li and Be abundances in M 67 together as a function of
decreasing temperature on the same vertical scale in Figure 9.  It is
clear that Li declines with decreasing temperature more steeply than
Be does.  The hottest stars, nearest the turn-off, are only moderately
depleted from the initial (i.e.~meteoritic) values.  The depletion
increases as these subgiants evolve toward cooler temperatures.

This span in A(Be) and A(Li) with temperature is shown in Figures 10
and 11.  The Be abundance, shown in Figure 10, drops from A(Be) = 1.2
to $-$0.5, a factor of 50, over the drop of 1000 K in temperature, or
between 1.26 to 1.32 M$_{\odot}$ in stellar mass.  The lowest-mass,
least-evolved stars, have the most Be, i.e.~the least Be depletion.
For A(Li), shown in Figure 11, the drop is a factor of 400 over the
drop of 1300 K in temperature in the same mass range.  This is a
reflection of the larger drop in Li than in Be as found in the
main-sequence Li-Be dip, plus a potential drop due to
post-main-sequence dilution.

In these two figures we show the model predictions of Sills \&
Deliyannis (2000) for Be and Li in M 67 subgiants caused by
rotationally-induced mixing.  The trend of decline in the Be
depletions with temperature is off set by 500 K from the observations;
the observed depletions set in at a higher temperature.  Even the
hottest, least massive star, F3, has depleted Be with A(Be) = 1.17
compared to the ``initial'' value.  (That ``initial'' value for M
67 is somewhat unclear: Lodders (2003) give A(Be) = 1.41 from
meteorites while Asplund et al.~(2009) give 1.30 for meteorites from
Lodders et al.~(2009) and solar A(Be) as 1.38.  From Figure 2 we see
that the peak Be in the Hyades stars near 6000 K is 1.42 which is
presumably their ``initial'' Be abundance.)  For F3 the error in our
A(Be) is $\pm$0.09.

Inasmuch as we now have Be abundances, we are able to use both
elements to examine the Li and Be decline to understand the stellar
interior processes.  In Figure 12 we plot A(Li) vs.~A(Be).  The
left panel shows the main-sequence field and cluster stars from
Boesgaard et al.~(2004b) in the temperature range 6300 - 6650 K, with
the equation given in section 2.  That slope is 0.43 $\pm$0.04 and
results from main-sequence depletion of both Li and Be.  The right
panel is for our nine M 67 subgiants.  We have made a linear fit
through these points and find a slope of +0.65 $\pm$0.06.  This
indicates that the Li abundance is falling much faster than the Be
abundance in these subgiants.  Perhaps this is due to the greater
depletion of Li during the main-sequence life of this older cluster as
well as the effects of dilution.  The mechanism for depletion would be
more effective on Li than Be.  In addition there would be greater
effects of dilution on Li than on Be because the Li reservoir is
smaller that the Be reservoir.

Models that include rotationally-induced mixing for Li and Be in M 67
subgiants from Sills \& Deliyannis (2000) are shown Figure 13.  This
reveals a more complex relationship than the simple linear fit in
Figure 11.  And it is a seemingly better fit to the observations.  If
other mechanisms also play a role, they must do so in a way that does
not alter the Li/Be ratio significantly. Sills \& Deliyannis (2000)
also discuss the predicted effects of mass loss and diffusion on M 67
subgiant Li and Be abundances.  Our data clearly rule out both as the
dominant Li- and Be-depleting mechanism.

We have tried to place the two subgiants observed for Be by Randich et
al.~(2007) on the same (Pace) scale.  The temperature differences are
small: $-$32 K and +64 K.  However, it is not clear what value of log
g was used by them.  Assuming it is 4.1 we can calculate the change
from their value for A(Be).  We have not plotted these two points in
Figures 9, 11, and 12 due to the issue with log g.  The data for the
two stars are consistent with our results.  These values would then
become: Y1248 = S1039: T = 5937, A(Be) = $-$0.08, A(Li) = 1.26 and Y923 =
S1239: T = 5541, A(Be) = $-$0.08, A(Li) = $<$0.08.  

\section{SUMMARY AND CONCLUSIONS}

Dramatic depletions of Li have been found in the subgiant stars in M
67 as summarized by Pace et al.~(2012).  Our results for Be produce
additional information to those for Li because they provide a probe
into a deeper layer of the stellar interior.  We have taken
high-resolution spectra with Keck I with HIRES of nine subgiants in M
67 to determine the abundances of Be.  We have used the set of stellar
parameters and Li abundances established by Pace et al.~in our analysis.

The subgiants in M 67 have evolved from stars that once were in the
Li-Be dip region that is found in younger clusters.  That dip is
located in a 400 K region centered around 6650 K.  Therefore, before
becoming subgiants, they would have depleted some Li and Be from their
initial amounts.  As they evolved from the main-sequence, the
deepening convection zone would begin to dilute the remaining Li and
Be.  The effects of both main-sequence depletion and
post-main-sequence dilution are less effective on Be than Li both
observationally and theoretically.

The star, F3, is at the turnoff mass of 1.26 M$_{\odot}$ and is
depleted in Li by a factor of 4 from the initial value and depleted in
Be by a factor of 1.8.  The evolutionary effects on Li and Be are seen
in the subgiants from the low-mass stars to the more evolved ones,
corresponding to masses from 1.26 to 1.32 M$_{\odot}$ and temperatures
from 6400 K to 5400 K.  We can see in Figure 9 that A(Li) drops by a
factor of 400, while A(Be) declines by a factor of about 50 over that
temperature and mass range.  The decline in A(Li) is a steeper drop
with temperature than the decline in A(Be).

This greater decline in A(Li) is also shown in Figure 12 where the
slope between A(Li) and A(Be) is 0.68 $\pm$0.06.  That slope is
steeper than that main-sequence stars on the cool side of the Li-Be
dip which is 0.43 $\pm$0.04.  Again this shows the greater susceptibility of
Li, with respect to Be, for both main-sequence depletion and
post-main-sequence dilution.

The models that include the effects of stellar rotation on the
internal mixing processes in stars of Sills \& Deliyannis (2000)
predict the observed decline in Li and Be in the subgiants.  It is the
relative abundances of Li and Be that give the most powerful insights
about the physical processes occurring inside stars.  So, importantly,
those models produce an excellent fit for the striking relationship
between A(Li) amd A(Be) shown in Figure 13.

Both Figures 7 and 8 show that the value for solar A(Li) fits well
with the observed range in A(Li) in M 67 stars of 1 solar mass and
solar temperature.  The Sun appears to have a normal Li content in
spite of the prediction from standard solar models that the Sun should
be depleted by only a factor of 3 rather than the observed factor of
200.

\acknowledgements We are grateful to the Keck Observatory support
astronomers for their knowledgeable assistance during our observing
runs.  We appreciate the observing help on the November, 2017 run by
Ashley Chontos.  CPD acknowledges support from the NSF through grant
AST-1211699.

\clearpage
\begin{deluxetable}{rrrrcclcc}

\tablenum{1}
\tablewidth{0pc}
\tablecaption{Log of the Keck/HIRES Be Observations in M 67 Subgiants}
\tablehead{
\colhead{F} & \colhead{Sand} & \colhead{MMJ} & \colhead{YBP} &
\colhead{V\tablenotemark{1}} & \colhead{B-V\tablenotemark{1}} & 
\colhead{(Date-UT)} &
\colhead{Exp(min)} & \colhead{S/N} } 
\startdata 
3 & 610 & 5042 & \nodata & 12.856 & 0.531 & 2014 Jan 16 & 120 & 44 \\ 
46  & 806  & 5350 & 1632 & 12.782 & 0.813 & 2014 Dec 27 & 120 & \nodata \\ 
    &      &      &      &        &       & 2017 Nov 10 & \phn60 & 35 \\
127 & 995  & 5675 & 963  & 12.755 & 0.559 & 2014 Dec 28 & 135 &\nodata  \\ 
    &      &      &      &        &       & 2017 Nov 10 & \phn60 & 45 \\
182 & 2207 & 5929 & 1070 & 12.631 & 0.620 & 2017 Nov 10 & \phn30 & \nodata \\
    &      &      &      &        &       & 2017 Nov 11 & 120    & 47 \\
202 & 1275 & 6018 & 1320 & 12.562 & 0.593 & 2014 Dec 27 & 120 & 47 \\ 
210 & 1273 & 6047 & 1318 & 12.219 & 0.567 & 2014 Jan 16 & \phn60 & \nodata \\
    &      &      &      &        &       & 2017 Nov 11 & \phn30 & 37 \\ 
243 & 1268 & 6177 & 1258 & 12.617 & 0.581 & 2014 Jan 16 & 120 & 49 \\ 
272 & 1487 &\nodata&1876 & 12.577 & 0.641 & 2014 Dec 27 & \phn90 & \nodata \\
    &      &      &      &        &       & 2014 Dec 28 & \phn45 & 47 \\ \
289 & 1607 &\nodata&\nodata & 12.620 & 0.560 & 2014 Dec 28 & \phn45 & 24 \\ 
\enddata
\tablenotetext{1} {V and B-V values are from CCD photometry by Montogomery et
al.~(1993) except for Sand1607 which is from Sanders (1977) and YBP 1876 
which is  from Yadav et al.~(2008).}
\end{deluxetable}



\clearpage
\begin{deluxetable}{rrcrccc}

\tablenum{2} 
\tablewidth{0pc} 
\tablecaption{Stellar Parameters}
\tablehead{ 
\colhead{F} & \colhead{Sand} & \colhead{MMJ} &
\colhead{YBP} & \colhead{$T_{\rm eff}$(K)\tablenotemark{1}} &
\colhead{log g}(K)\tablenotemark{1} & \colhead{$\xi$} 
} 
\startdata 
3   & 610  & 5042 & \nodata & 6393 & 4.000 & 2.21 \\ 
46  & 806  & 5350 & 1632 & 5398 & 3.717 & 1.79 \\ 
127 & 995  & 5675 &  963 & 6106 & 4.002 & 1.98 \\
182 & 2207 & 5929 & 1070 & 5920 & 3.918 & 1.94 \\ 
202 & 1275 & 6018 & 1320 & 5967 & 3.884 & 2.02 \\ 
210 & 1273 & 6047 & 1318 & 6082 & 3.831 & 2.19 \\ 
243 & 1268 & 6177 & 1258 & 6031 & 3.911 & 2.04 \\ 
272 & 1487 & \nodata & 1876 & 5852 & 3.791 & 2.05 \\ 
289 & 1607 & \nodata &\nodata & 6124 & 3.919 & 2.01 \\ 
\enddata
\tablenotetext{1} {These values are from Pace et al.~(2012)}
\end{deluxetable}


\clearpage
\begin{deluxetable}{rlcccr}

\tablenum{3}
\tablewidth{0pc}
\tablecaption{Abundace Results}
\tablehead{
\colhead{F} & \colhead{Name} & \colhead{$T_{\rm eff}$(K)\tablenotemark{1}} 
& \colhead{Mass (M$_\odot$)\tablenotemark{1}}  & \colhead{A(Li)\tablenotemark{1}} & \colhead{A(Be)} 
}
\startdata
3    & S610  &  6393 & 1.26 & 2.72 & 1.17 \\
46   & Y1632 &  5398 & 1.32 & 0.12 & $\leq$$-$0.50 \\
127  & Y963  &  6106 & 1.27 & 2.03 & 1.10 \\
182  & Y1070 &  5920 & 1.28 & 1.20 & 0.30 \\
202  & Y1320 & 	5967 & 1.29 & 2.15 & 0.94 \\
210  & Y1318 &  6082 & 1.30 & 1.85 & 0.71 \\
243  & Y1258 &  6031 & 1.28 & 1.03 & 0.34 \\
272  & Y1876 &  5852 & 1.31 & 1.06 & 0.06 \\
289  & S1607 &  6124 & 1.28 & 1.71 & 0.74 \\
\enddata
\tablenotetext{1} {These values are from Pace et al.~(2012)}
\end{deluxetable}

\clearpage

\begin{figure}
\plotone{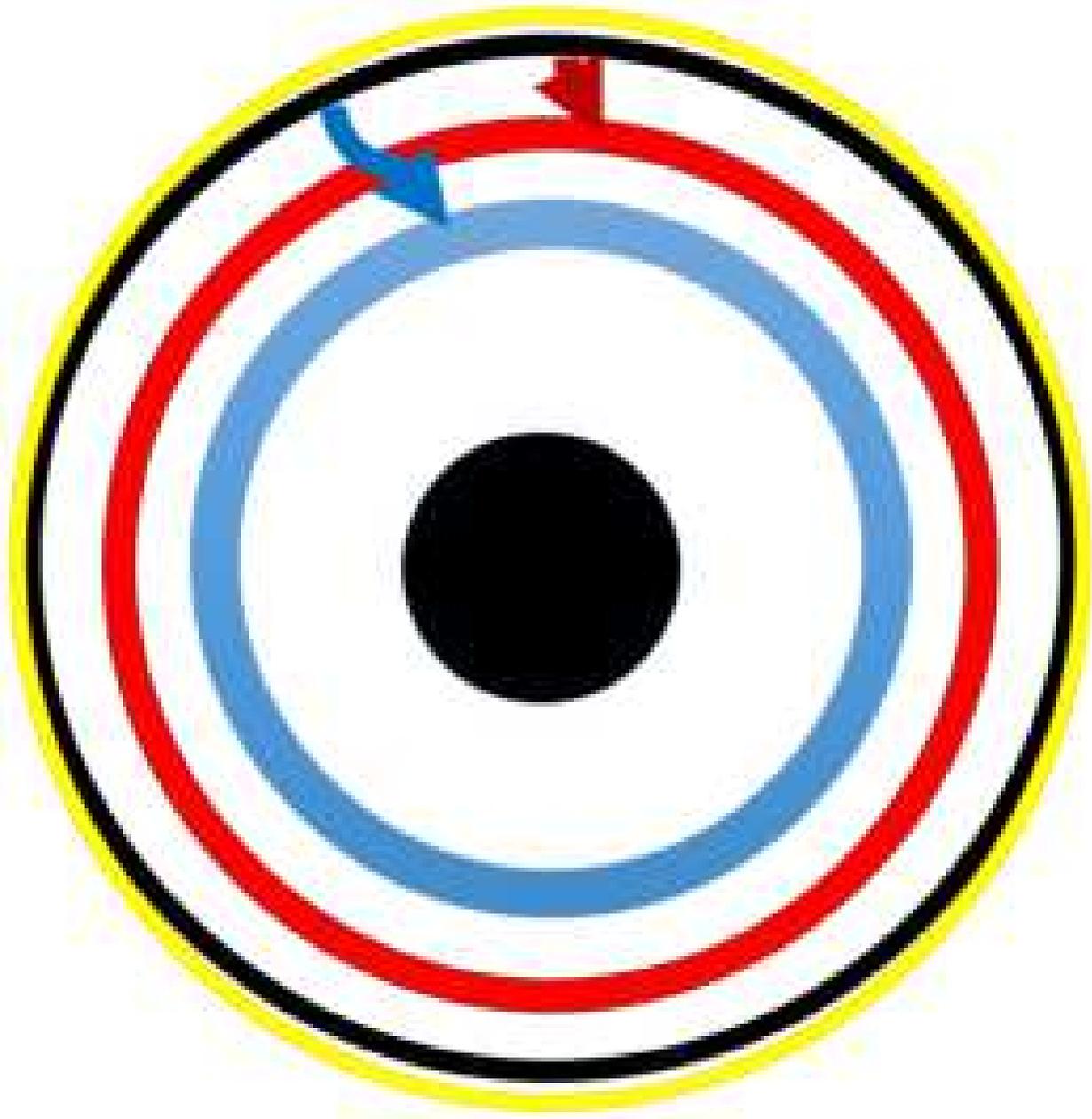}
\caption{A schematic model of a cross-section of the Sun.  The
photosphere is the outer yellow ring; the outer black ring represents
the surface convection zone.  All Li nuclei inside the red circle are
destroyed by nuclear reactions.  The arrows indicate the inward
mixing that that is required for the surface atoms of Li (red) and Be
(blue) to be destroyed.  Similarly, all Be nuclei inside the blue
circle destroyed by nuclear reactions.  This leaves regions in the
stellar interior that are devoid of Li and of Be.  The black core
indicates the region where there are energy-producing thermonuclear
reactions.  The red arrow indicates that Li atoms need to be
transported down to the region where they will be destroyed and the
blue arrow indicates the similar transport for Be atoms.}
\end{figure}

\clearpage

\epsscale{1.0}
\begin{figure}
\plotone{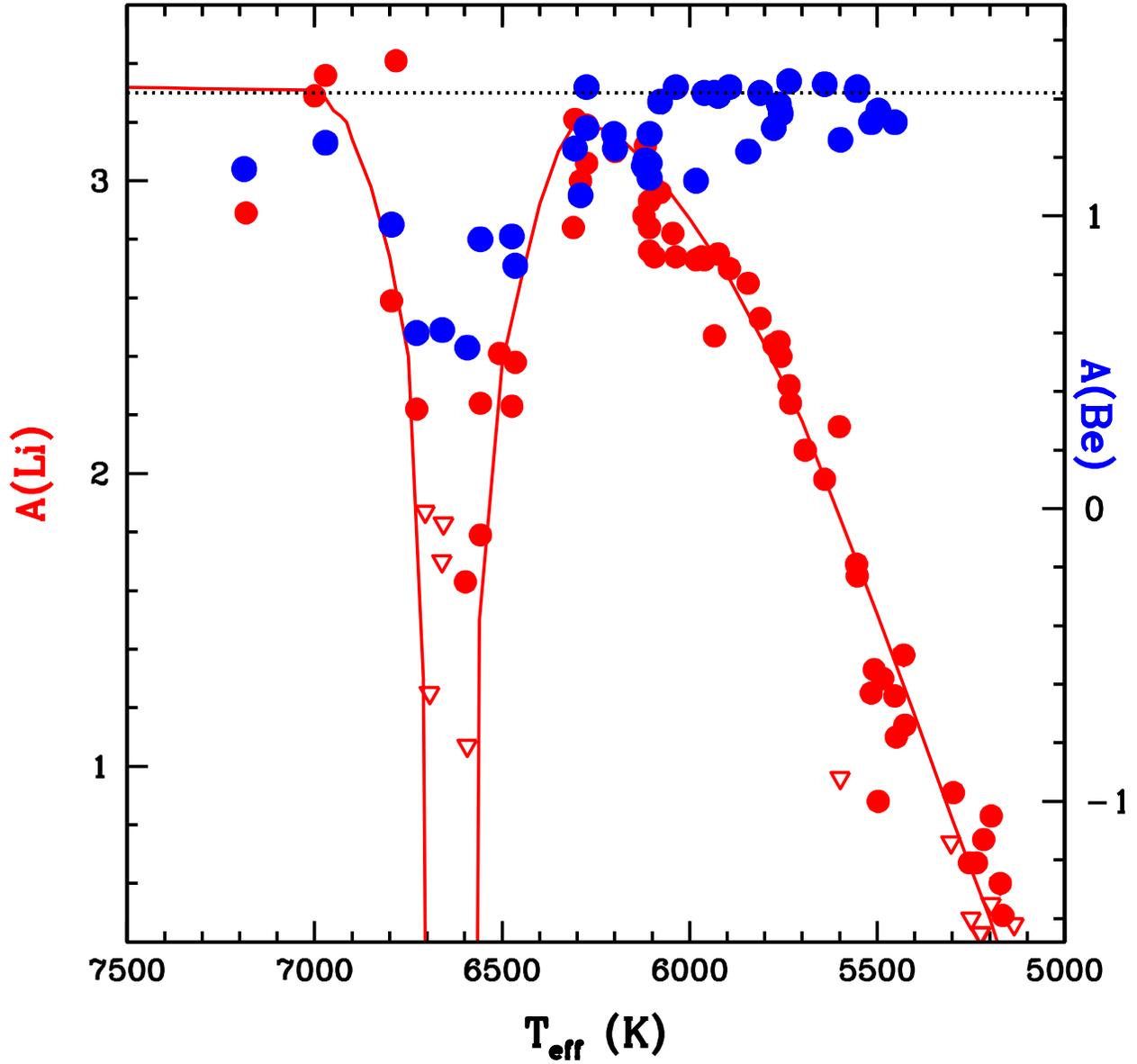}
\caption{Li and Be abundances in the Hyades on the same scale and
normalized to their respective solar system values. The left y-axis
shows A(Li) while the right y-axis shows A(Be).  The abundances are
taken from Boesgaard et al.~(2016).  The red solid line represents a
fit through the Li-temperature data}
\end{figure}

\clearpage
\epsscale{1.0}
\begin{figure}
\plotone{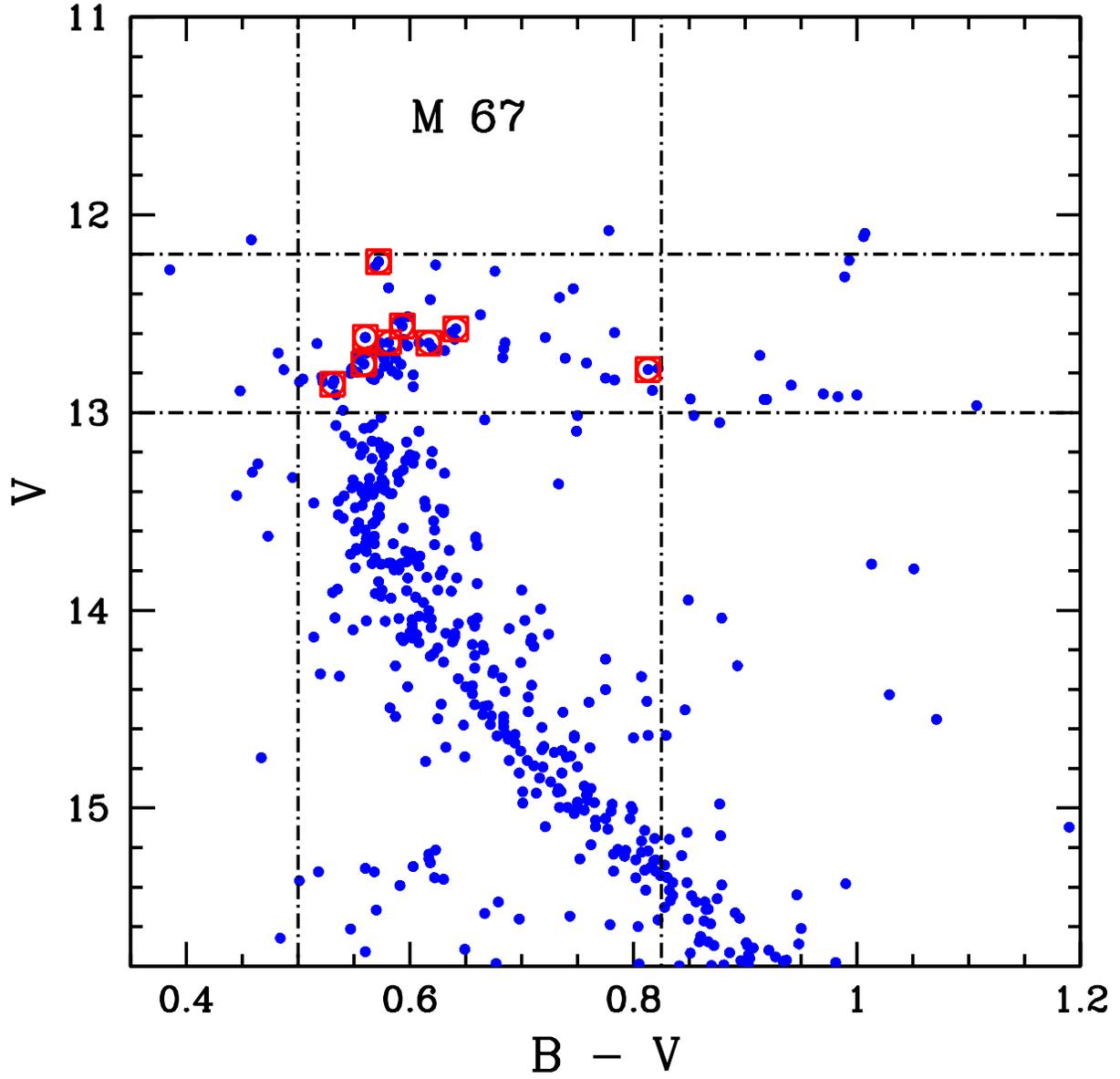}
\caption{Color-magnitude diagram of M 67.  The photometry is from Montgomery
et al.~(1993).  The vertical and horizontal lines enclose the region of our
target stars.  The red squares indicate the stars observed for Be.}
\end{figure}
\clearpage

\begin{figure}
\plotone{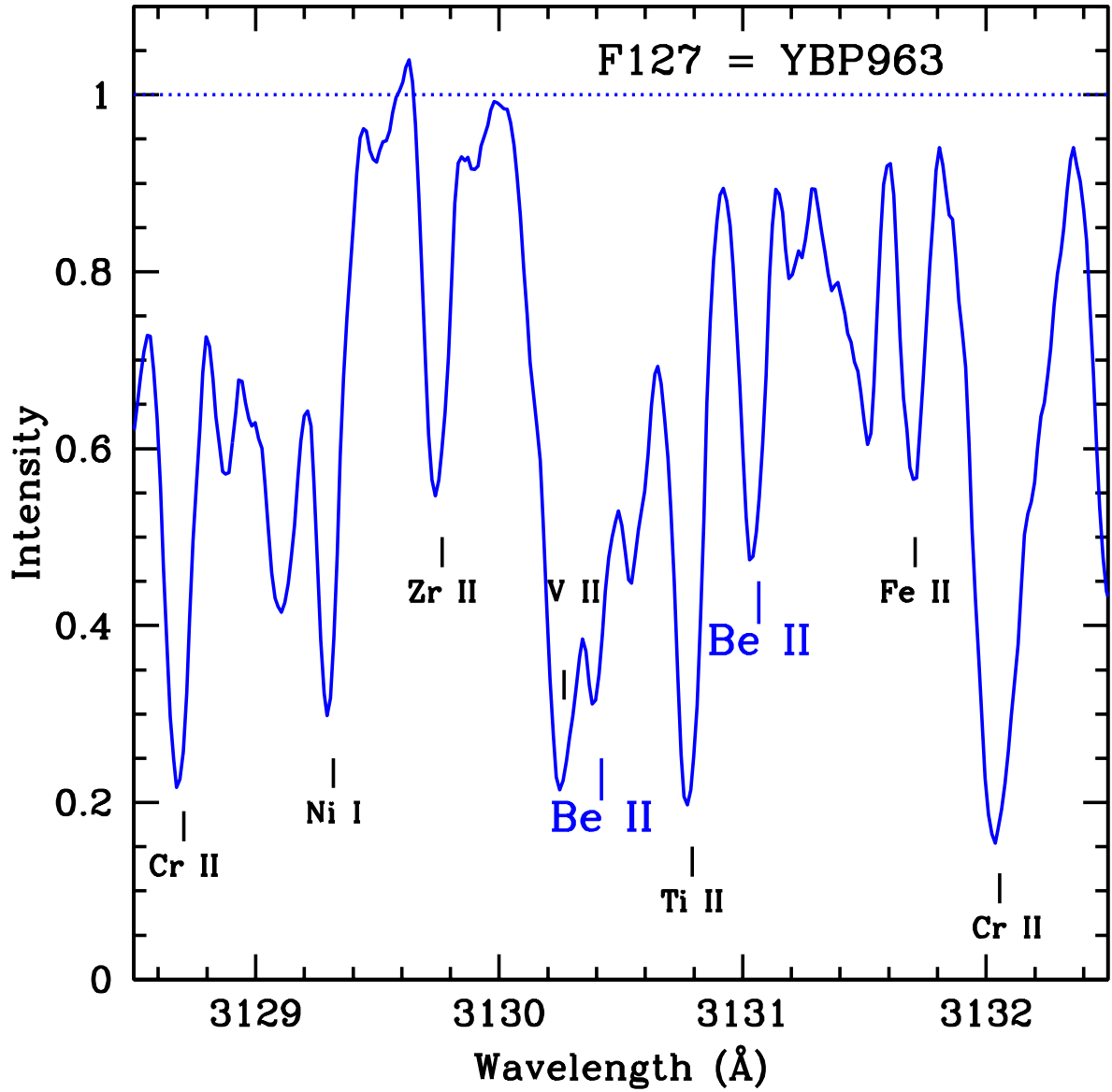}
\caption{The spectrum of F127 in a 4 \AA{} region surrounding the Be II
lines.  Some of the dominant lines in the region are identified.  All the
lines are blended with atomic and molecular features.}
\end{figure}

\begin{figure}
\plotone{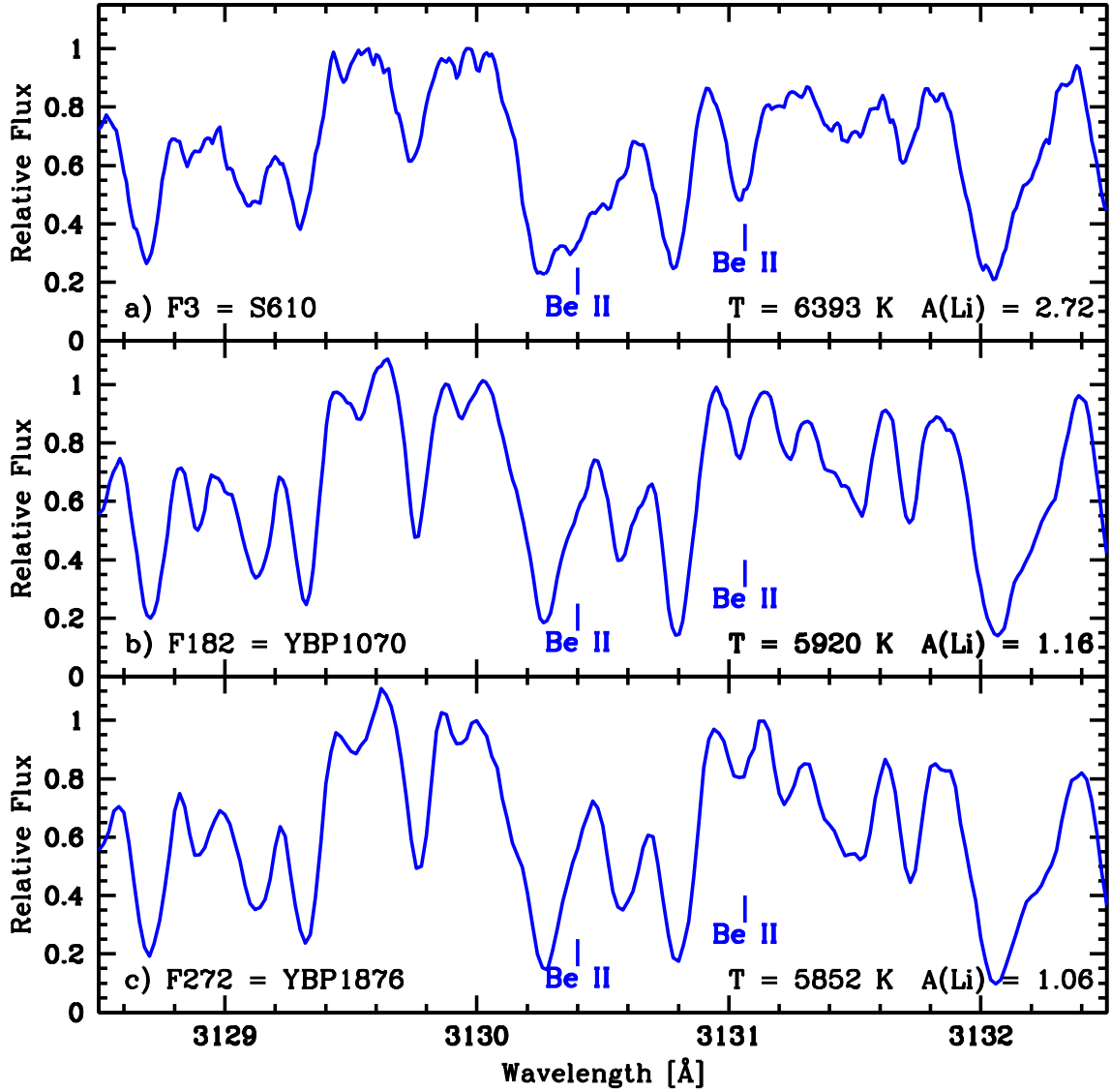}
\caption{Examples of the reduced spectra.  The positions of the Be II
lines are indicated.  These samples cover a range of temperatures and
Li abundances in our target stars.  The Be II lines are clearly weaker
in the low-Li stars in panels b) and c).}
\end{figure}

\begin{figure}
\plotone{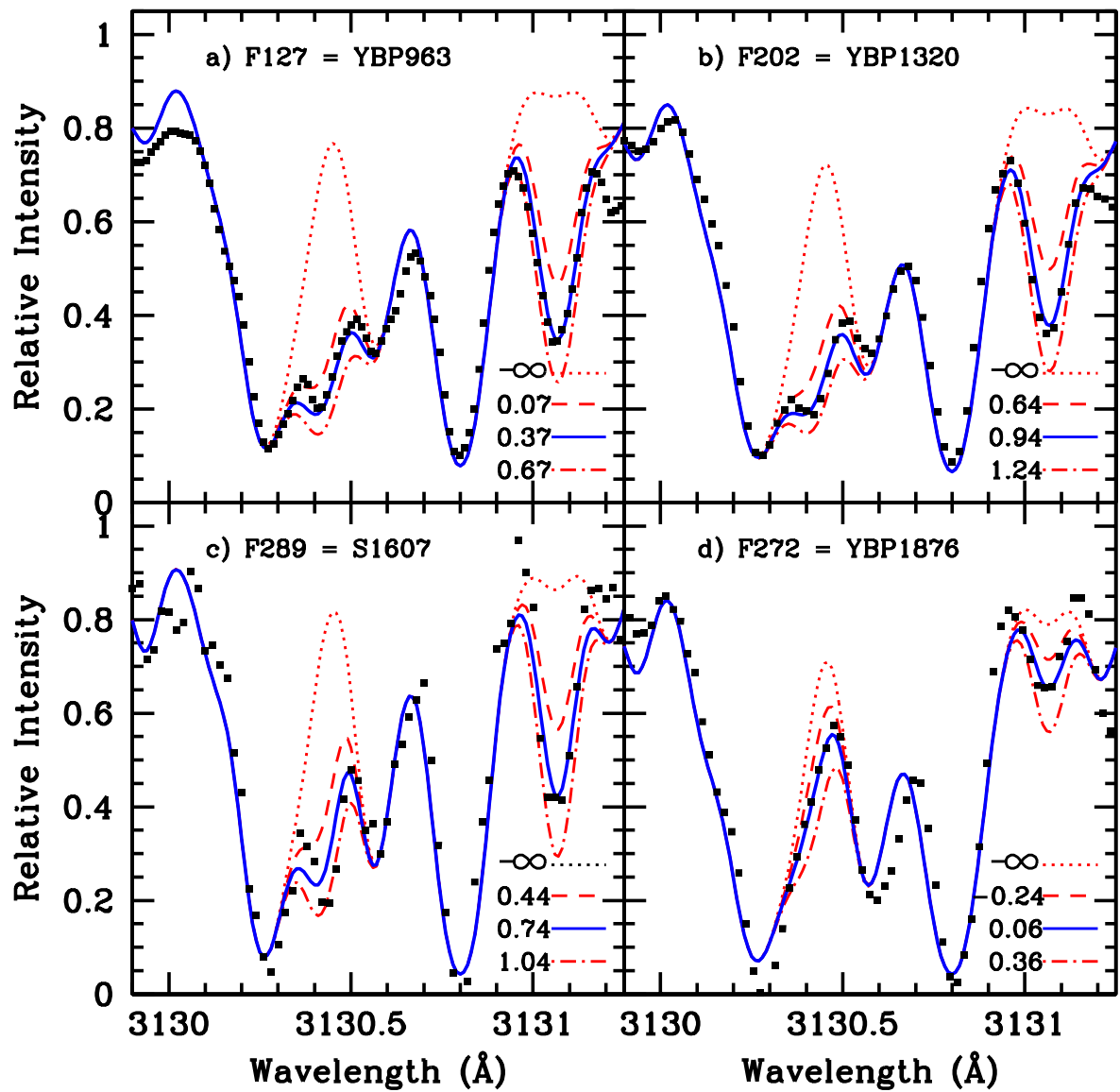}
\caption{Our spectrum syntheses of Be for four stars in our sample.  The
observed spectra are shown by the black dots.  The best fit is the blue solid
line.  The red dotted line corresponds to a synthesis with no Be at all.  The
red dashed line is a factor of two less Be and the red dot-dashed is for a
factor of two more Be.  The A(Be) values are indicated by the legend in the
lower right corner of each panel.  The high-mass, evolved star,
F272, clearly shows large depletion in Be as seen in panel d.}
\end{figure}

\clearpage

\epsscale{1.0}
\begin{figure}
\plotone{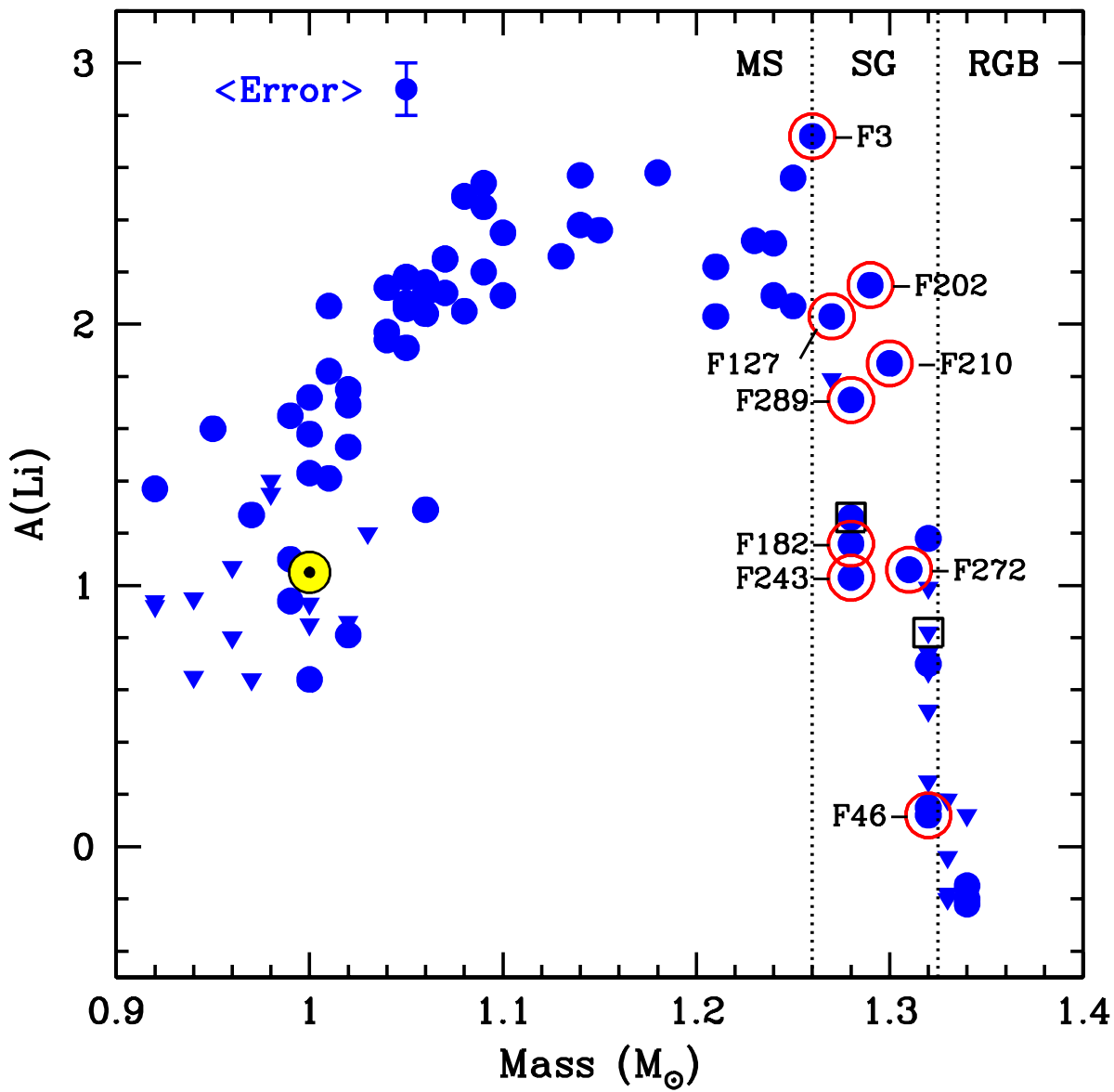}
\caption{The distribution of A(Li) with stellar mass from the values
of Pace et al.~(2012).  The inverted triangles indicate upper
limits on A(Li). The stars we have observed for Be are circled in
red.  They cover the range in A(Li) and mass.  The stellar
identifications (Fagerholm numbers) are shown for each star.  The
vertical dotted lines separate the main-sequence stars (MS) from the
subgiants (SG) and the giants (RGB).  The more massive subgiants are
the most evolved.  The two subgiants observed for Be by Randich et
al.~(2007) are indicated by the black squares.  The position for the
Sun is the yellow circle.}
\end{figure}

\clearpage

\begin{figure}
\plotone{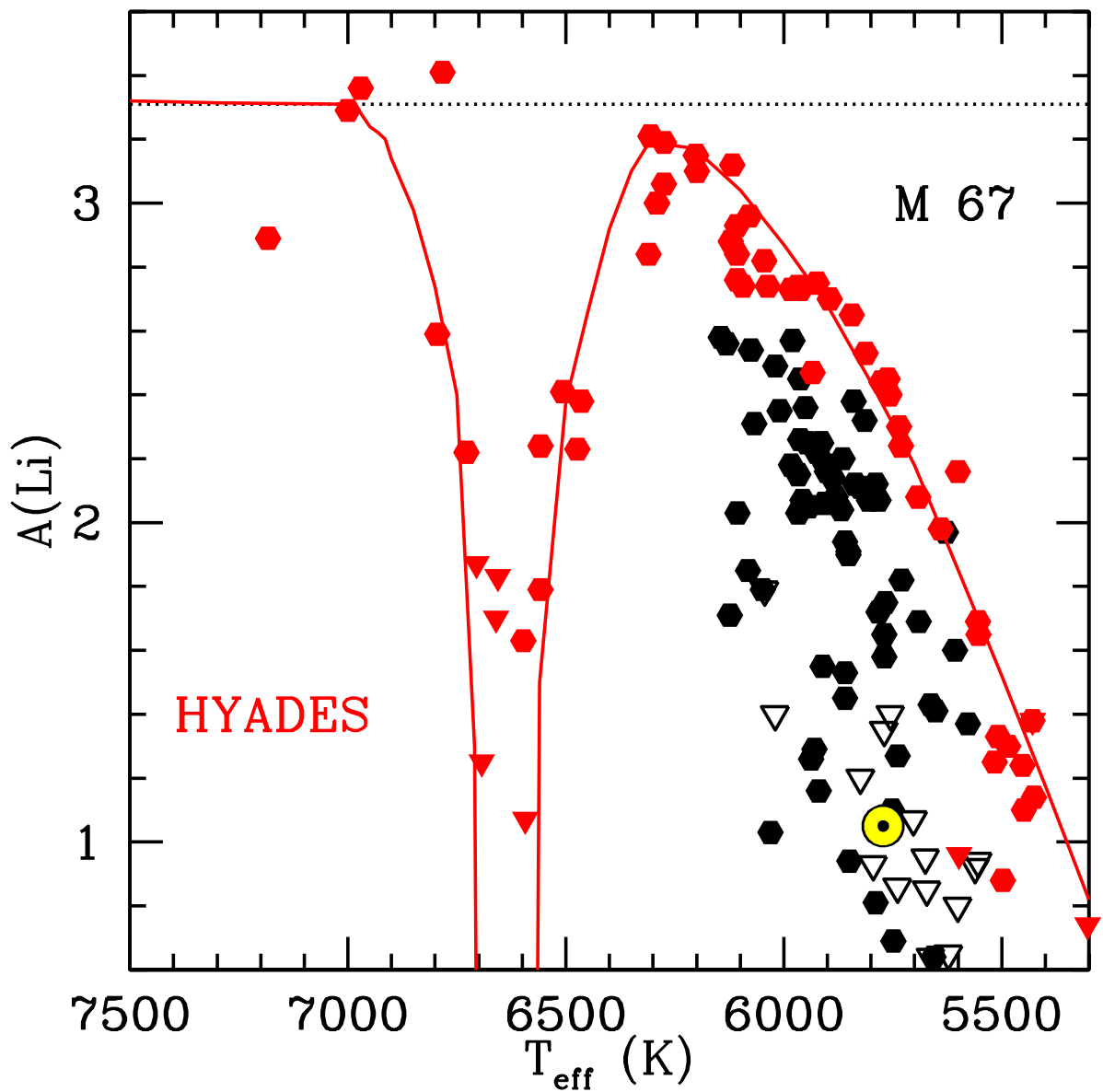}
\caption{The plot of A(Li) vs.~$T_{\rm eff}$ for the Hyades and for
the main-sequence stars (only) in M 67.  The inverted triangles
indicate upper limits on A(Li).  The M 67 main-sequence stars are only
on the lower part of the main sequence.  The stars that had inhabited
the Li-Be dip region are now subgiants and beyond.  One can see that
there is a much larger spread in A(Li) at a given temperature below
6300 K in M 67 than in the Hyades.  This is likely to be a result of
age and rotational spin-down effects.  The position for the Sun is the
yellow circle.}
\end{figure}

\clearpage

\begin{figure}
\plotone{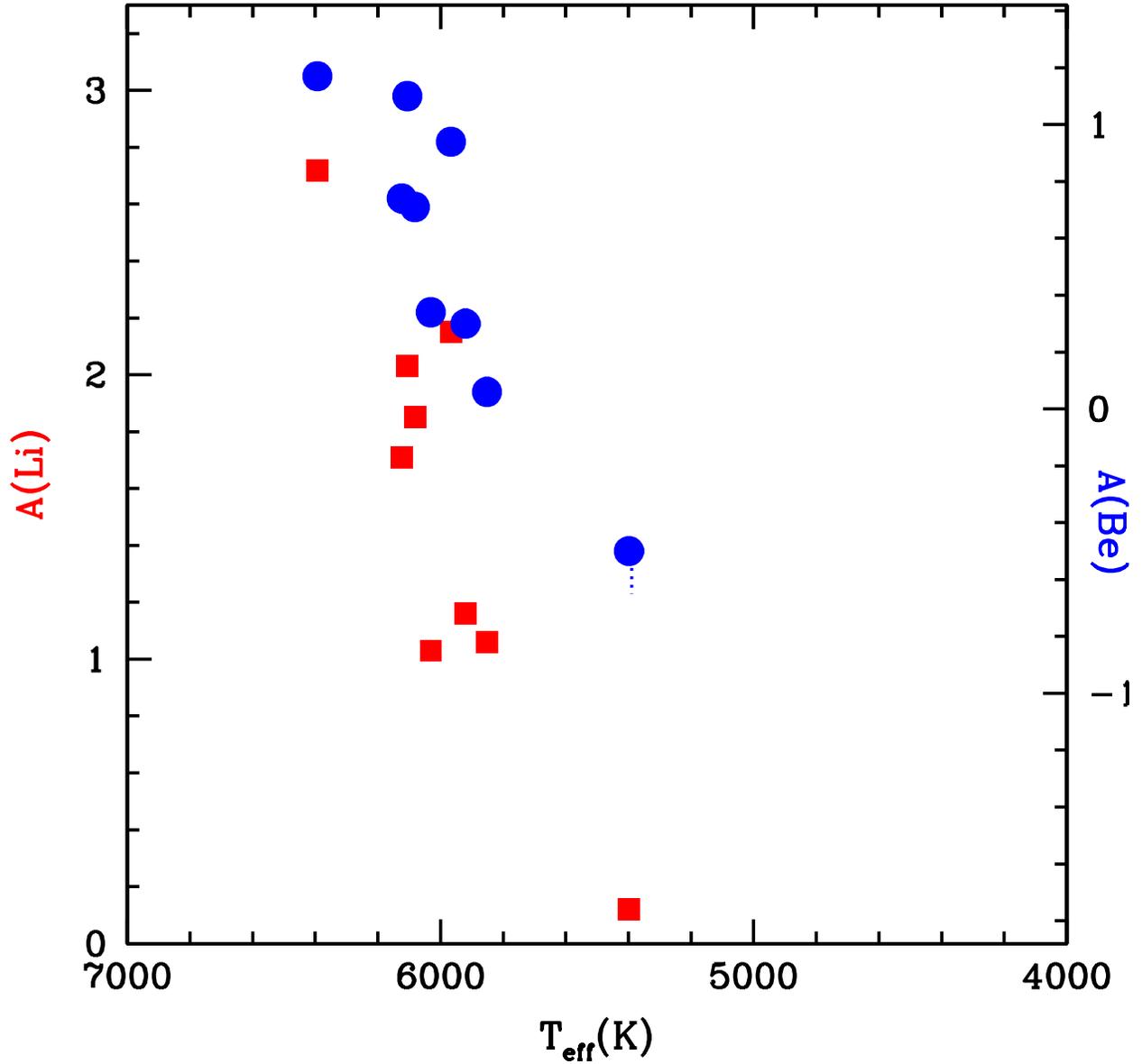}
\caption{Li and Be abundances in our M 67 subgiants on the same scale
and normalized to their respective solar system values. The left
y-axis shows A(Li) while the right y-axis shows A(Be).  The dotted
line emanating from the lowest A(Be) point (in this figure and
subsequent ones) is meant to signify the Be abundance is less than or
equal to that value of $-$0.50.  The Li abundance clearly drops faster
than the Be abundance.}
\end{figure}

\clearpage

\begin{figure}
\plotone{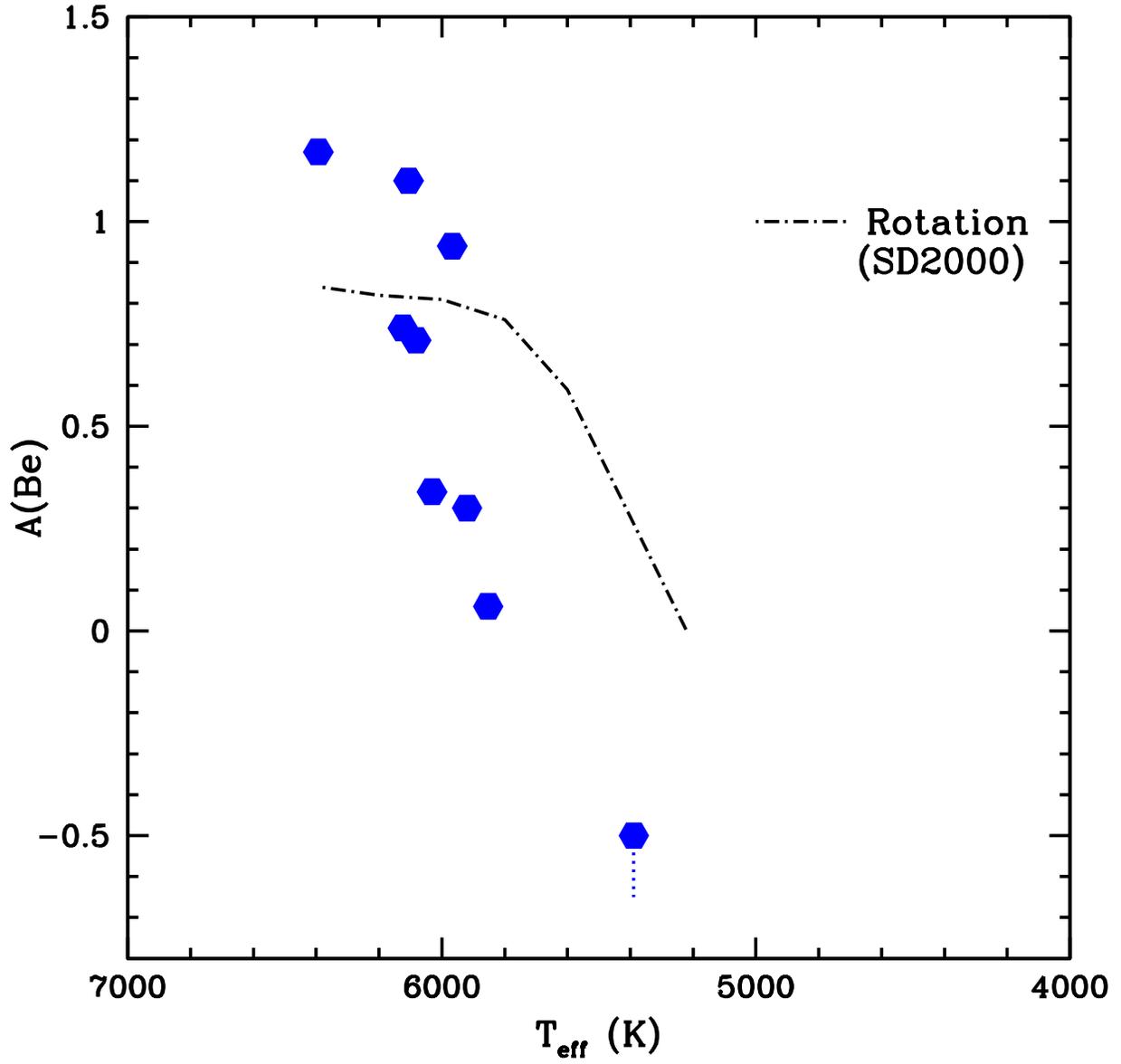}
\caption{Our Be abundances in the subgiants as a function of
temperature.  The dotted-dashed line show the model of Sills \& Deliyannis
(2000) for rotation effects on Be in M 67 subgiants.}
\end{figure}

\clearpage

\begin{figure}
\plotone{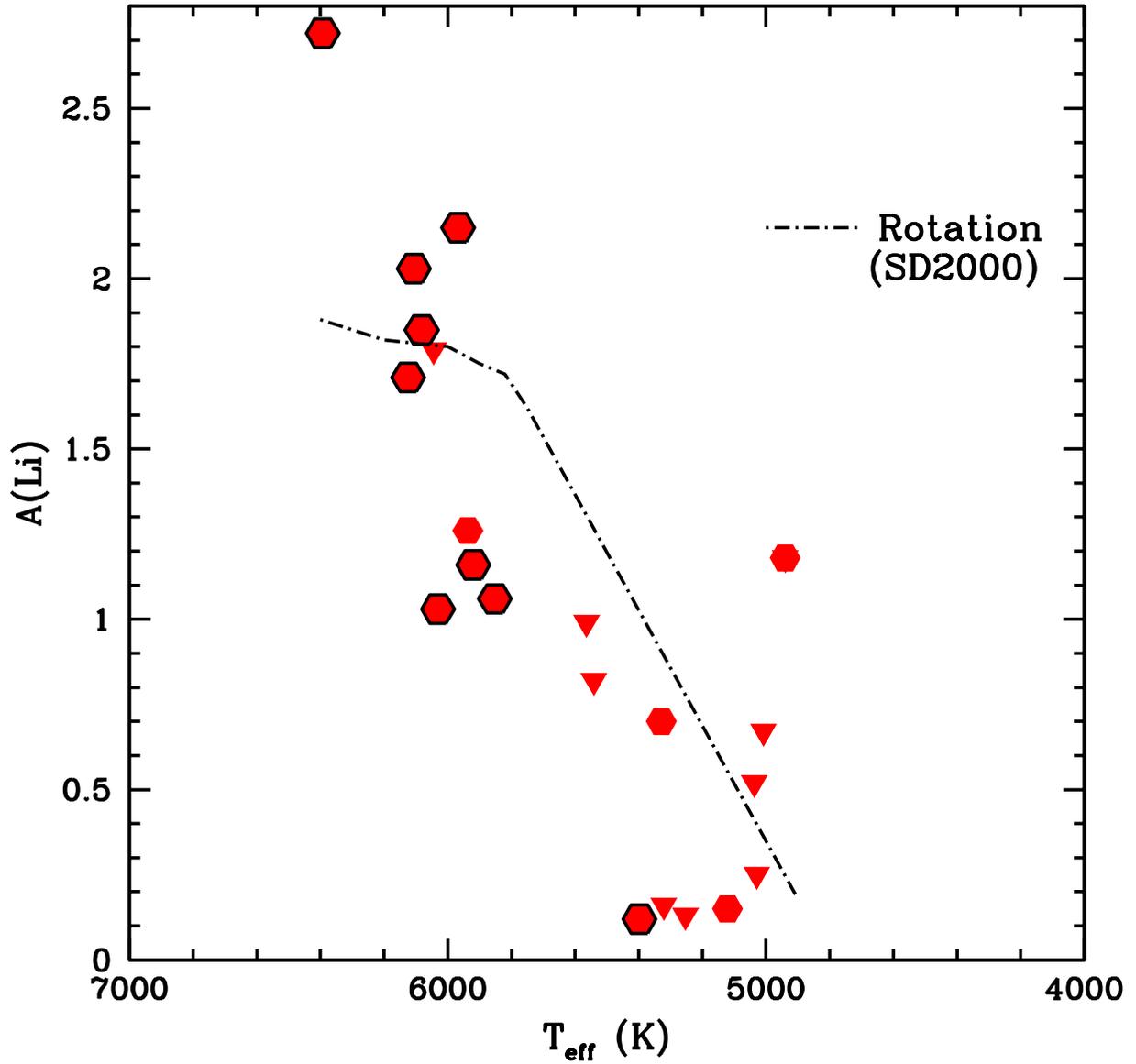}
\caption{The Li abundances from Pace et al.~(2012) vs.~temperature.
These Li data are only for stars with masses between 1.26 and 1.32
M$_{\odot}$.  The points outlined in black are those stars for which
we have Be abundances.  Again the dotted-dashed line is from the
calculations of Sills \& Deliyannis (2000) specifically for M 67
subgiants due to effects of stellar rotation.}
\end{figure}

\clearpage

\begin{figure}
\plottwo{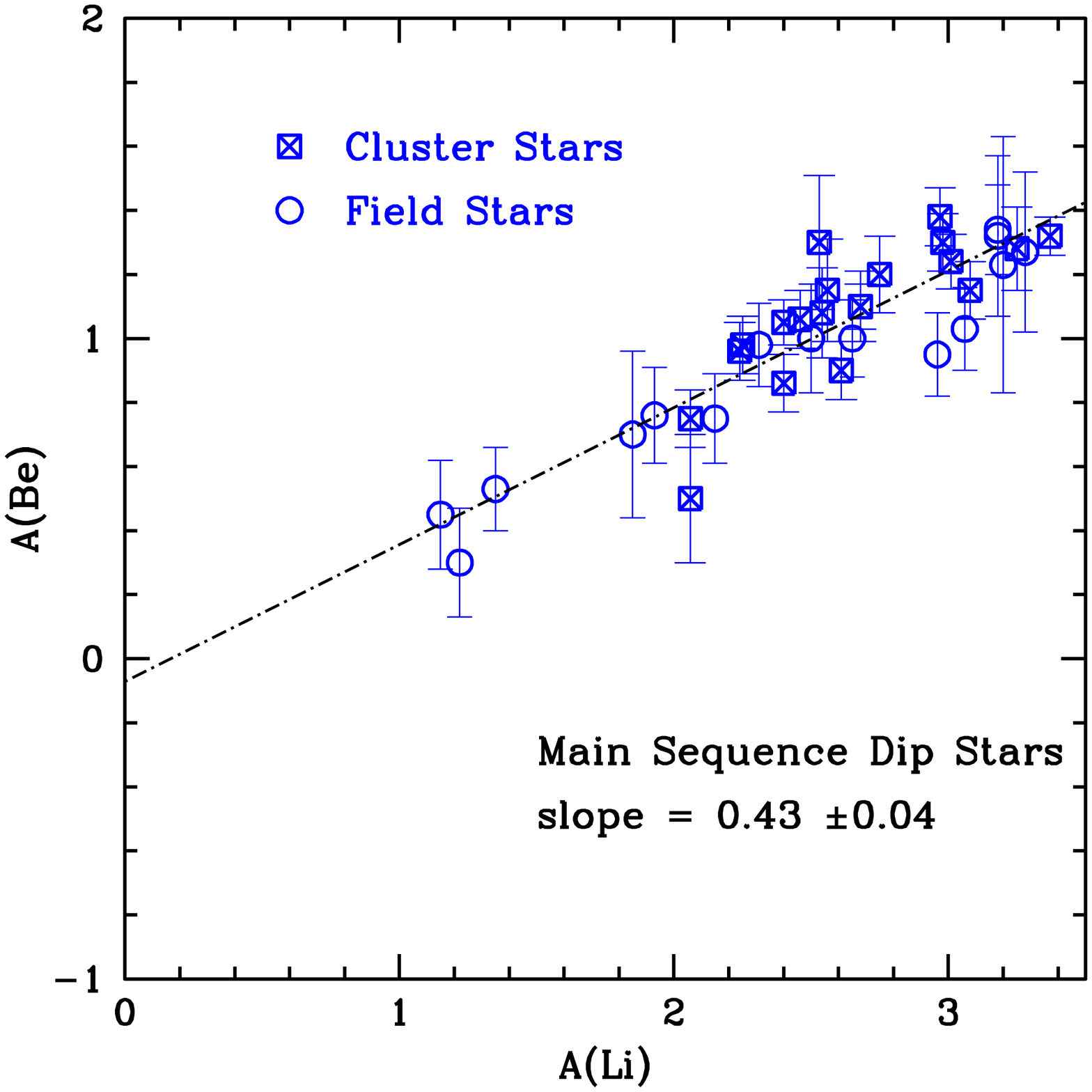}{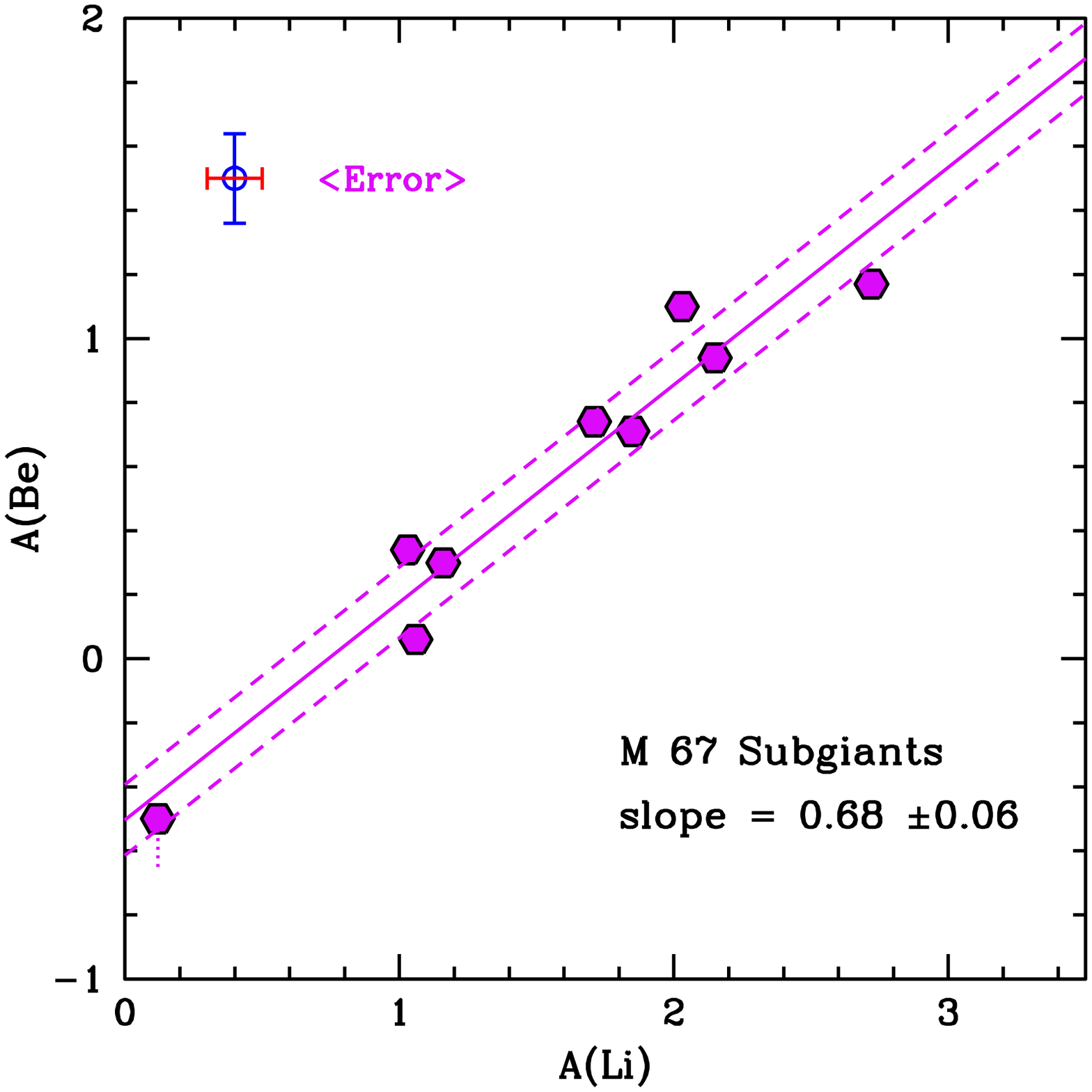}
\caption{The comparison of the abundances of Li and Be.  Left: The 35
field and cluster stars in the Li-Be dip from Boesgaard et
al.~(2004b).  The Li abundance falls off faster than that of Be in the
dip stars with a slope of 0.43.  Right: In the M 67 subgiants the Li
abundance falls off even faster relative to Be with a slope of 0.68.
As A(Li) goes down by a factor of 10, A(Be) is down by only 4.4.  That
steeper slope is the combined effect of both main-sequence depletion
and post-main-sequence dilution.  The solid line is the least squares
fit and the dashed lines show the 1 sigma errors.}
\end{figure}

\clearpage

\begin{figure}
\plotone{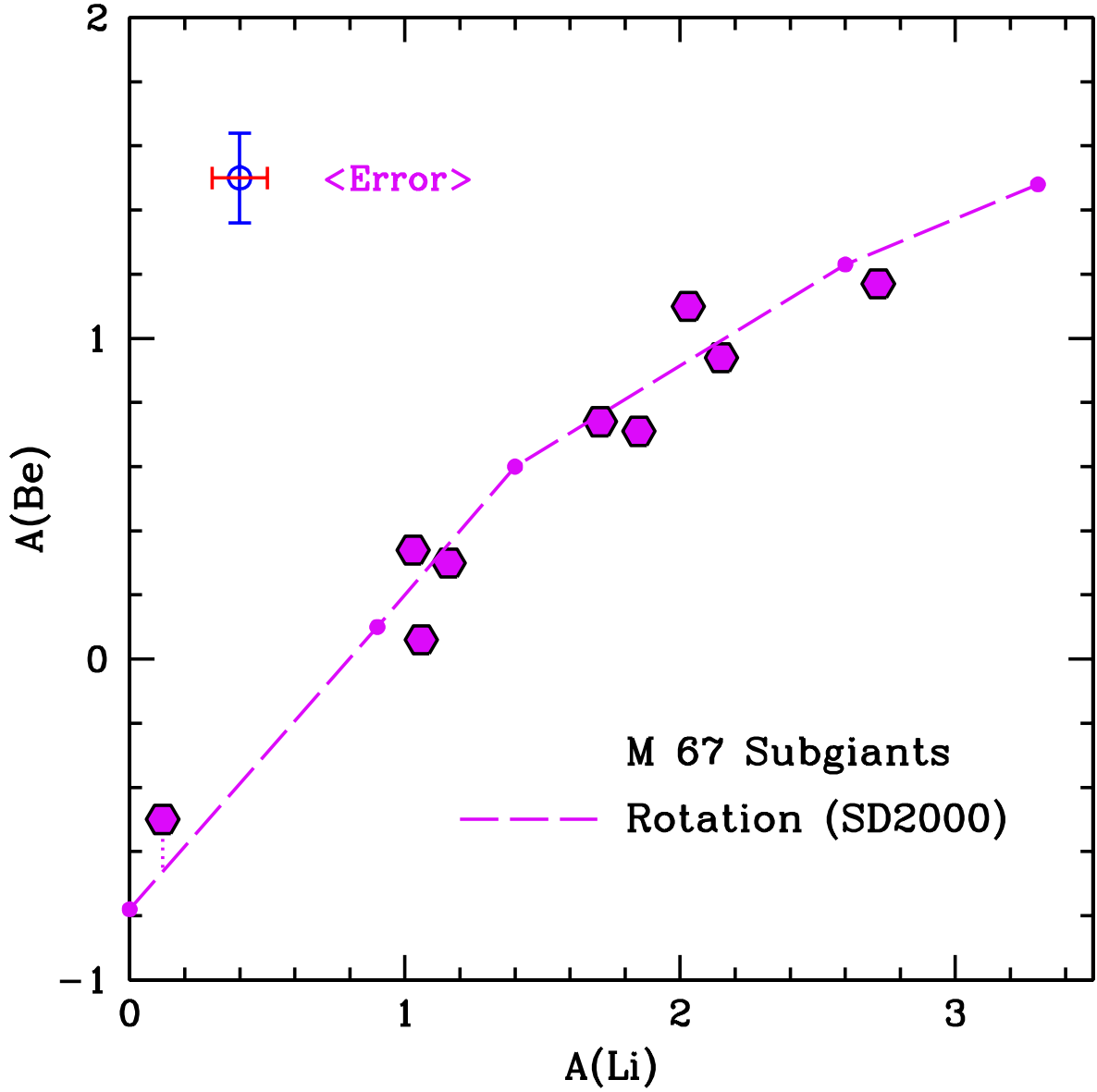}
\caption{The same data as in Figure 12b.  Here the curve corresponds to
the remarkably good fit to the data by the models of Sills \& Deliyannis 
which include rotation effects.}
\end{figure}

\clearpage

\end{document}